\documentclass[
    ,final            
  ]
  {aipproc}

\layoutstyle{6x9}

\def \pt	{p_{\perp}}
\def \mpt 	{\langle\pt\rangle}
\def \dphi	{\Delta\phi}

\begin{document}

\title{Supersonic Jets in Relativistic Heavy-Ion Collisions}



\author{Fuqiang Wang}{
address={Department of Physics, Purdue University, West Lafayette, Indiana 47907, USA}
}

\begin{abstract}
Mach-cone shock waves were proposed to explain the broad and perhaps double-peaked away-side 2-particle jet-correlations at RHIC; however, other mechanisms cannot be ruled out. Three-particle jet-correlation is needed in order to distinguish various physics mechanisms. In this talk the 3-particle jet-correlation measurements are presented and their implications are discussed.
\end{abstract}

\maketitle


\section{Introduction}

Jets and jet-correlations are good probes to study the medium created in relativistic heavy-ion collisions because their properties in vacuum can be calculated by perturbative quantum chromodynamics. Modifications to their properties in nuclear medium can be used to study the nature of the medium~\cite{Eloss}. While exclusive jet reconstruction is difficult in central heavy-ion collisions at RHIC, two-particle azimuthal correlations with a high transverse momentum ($\pt$) trigger particle have proved to be a powerful alternative~\cite{wp}.

\begin{figure}
\includegraphics[width=0.33\textwidth,height=0.25\textwidth,bbllx=130,bblly=160,bburx=455,bbury=375]{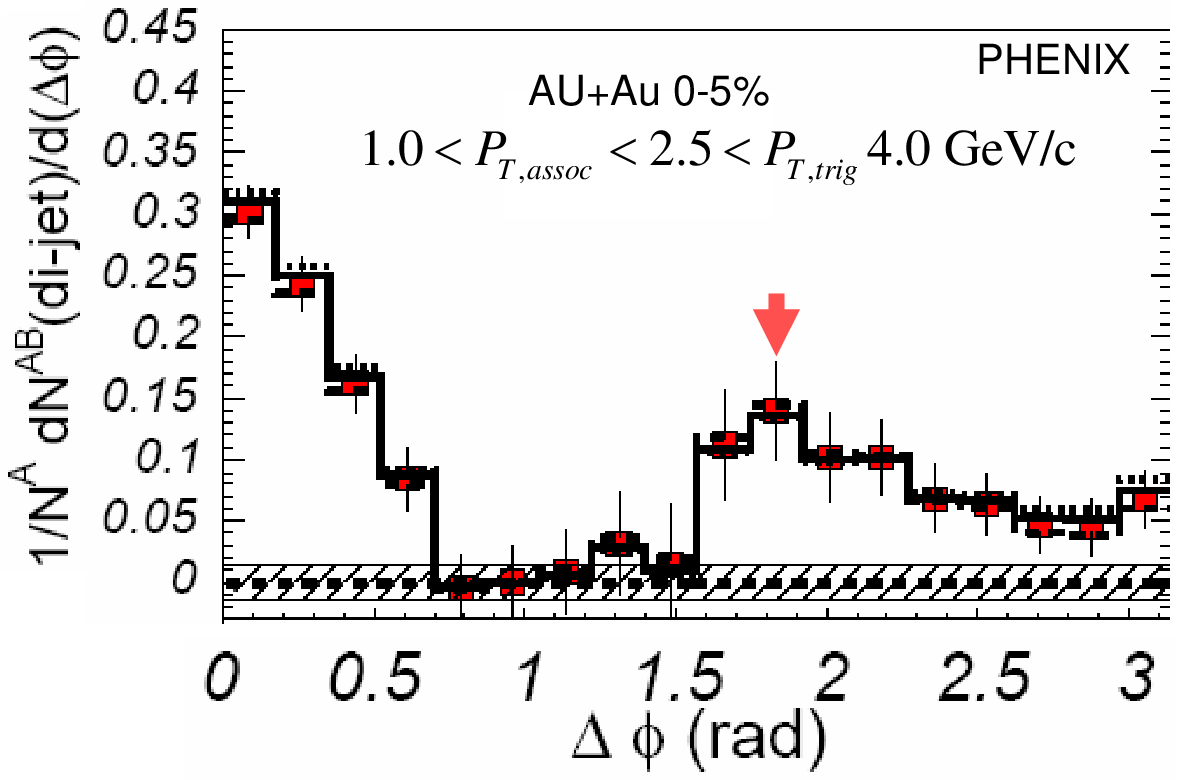}
\includegraphics[width=0.32\textwidth,height=0.235\textwidth,bbllx=20,bblly=0,bburx=565,bbury=400]{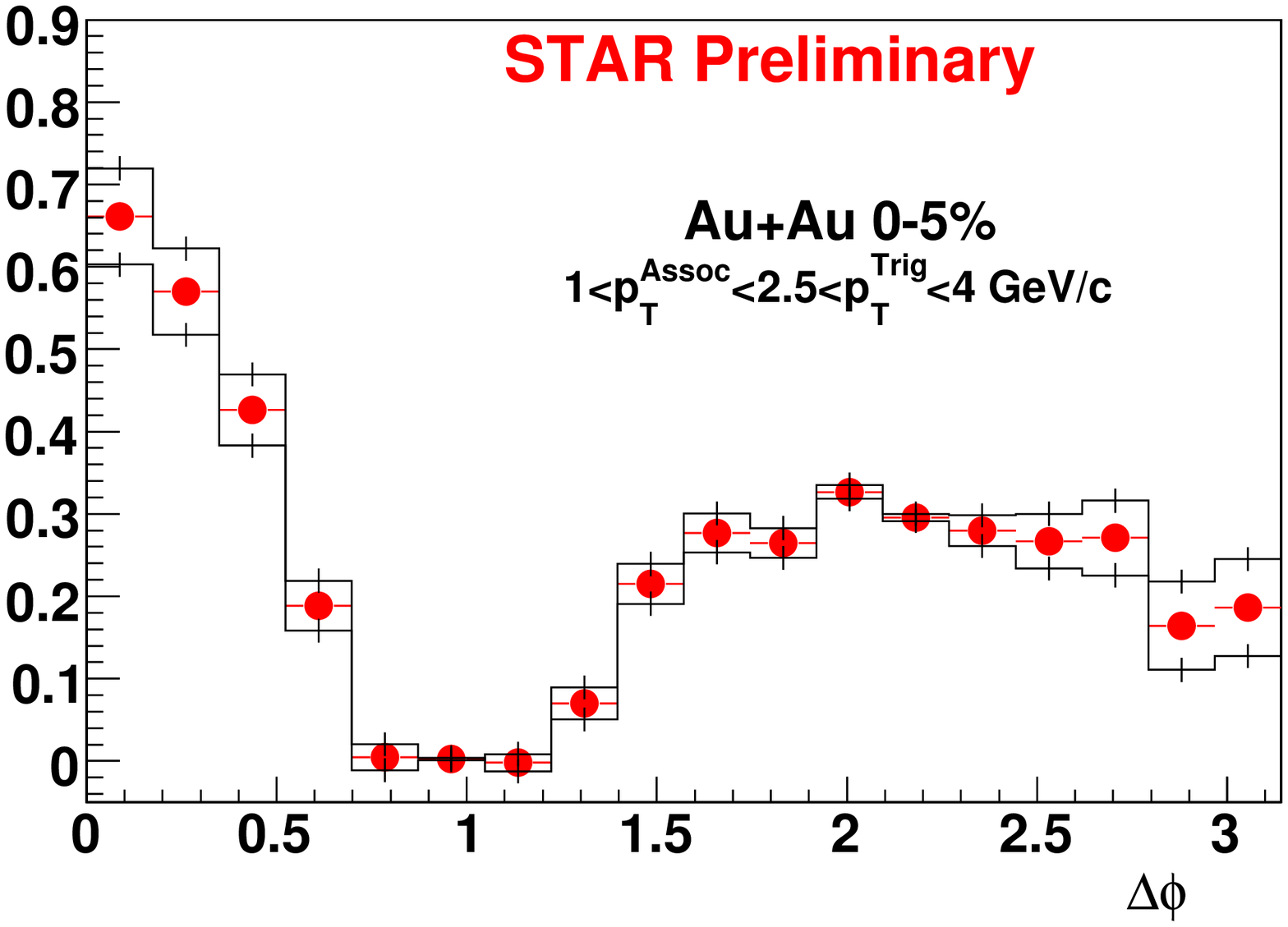}
\includegraphics[width=0.34\textwidth,height=0.24\textwidth,bbllx=12,bblly=20,bburx=500,bbury=328]{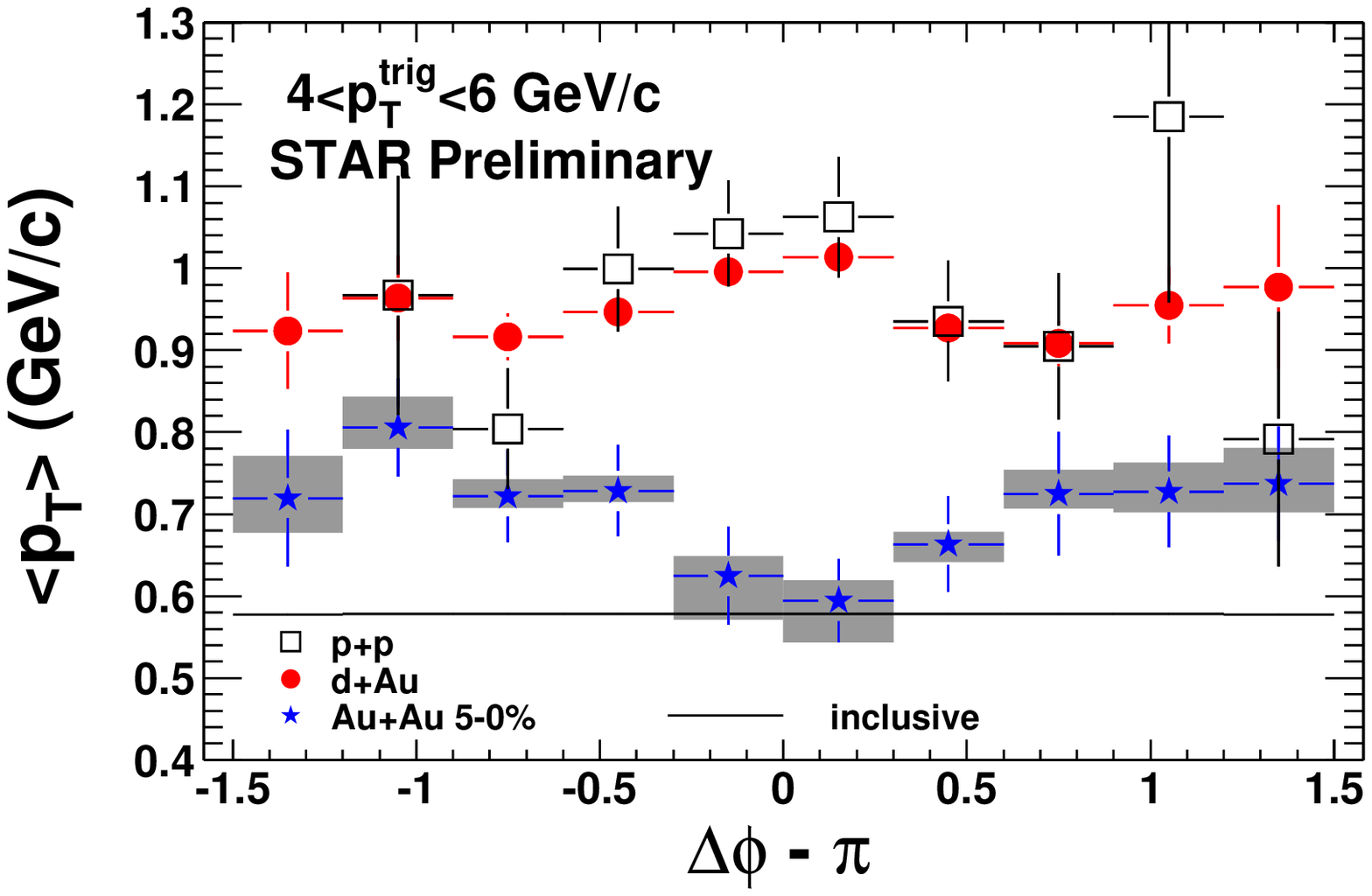}
\caption{(color online) Left panel: $\dphi$ correlations in central 5\% Au+Au collisions with $1<\pt<2.5<\pt^{trig}|<4$~GeV/$c$ from (a) PHENIX with $|\eta^{trig}|<0.35$ and $|\eta|<0.35$~\cite{dipPHENIX} and (b) STAR with $|\eta^{trig}<0.7$ and $|\eta|<1.0$~\cite{UleryQM05}. The histograms indicate systematic uncertainties. Right panel: Away-side correlated hadrons $\mpt$ versus $\dphi$ for $4<\pt^{trig}<6$~GeV/$c$ from STAR~\cite{UleryQM05}. Result for $3<\pt^{trig}<4$~GeV/$c$ is similar. Shaded areas indicate systematic uncertainties for the central Au+Au data.}
\label{fig1}
\end{figure}

The first study of azimuthal correlations between low $\pt$ hadrons and modest high $\pt$ particles has revealed rich information~\cite{jetspectra}. The correlated hadrons on the away side of the trigger particle are found to be broadly distributed; their energy distribution is similar to that of the bulk medium particles indicating partial equilibration. The away-side broadening becomes more prominent with lower trigger $\pt^{trig}$ and higher associated $\pt$; the azimuthal correlations with $1<\pt<2.5<\pt^{trig}<4$~GeV/$c$ are shown in Fig.~\ref{fig1} for central Au+Au collisions from PHENIX~\cite{dipPHENIX} (left panel) and STAR~\cite{awaympt,UleryQM05} (middle panel). Moreover, the average $\mpt$ of the away-side correlated hadrons shows a novel behavior as depicted in Fig.~\ref{fig1} (right panel) -- those more collimated with the trigger possess a lower $\mpt$~\cite{awaympt}, contrary to what is expected from jet fragmentation in vacuum. 

The broad distribution stimulated many theoretical investigations. In particular, Mach-cone shock waves were suggested as a possible physics mechanism~\cite{machcone} -- particles are emitted on a cone due to collective excitations, and their projection onto the azimuth  results in a double-peak structure. The generation of Mach-cone shock waves seems inevitable given that the medium is hydrodynamic~\cite{wp}, the jets are supersonic, and there are strong interactions between the jets and the medium~\cite{wp}. The Mach-cone angle is determined by the speed of sound of the medium and is independent of the associated particle $\pt$. Recently, $\check{\rm C}$erenkov gluon radiation was suggested as an alternative mechanism for conical emission~\cite{Cerenkov}; the cone angle in this case is dependent of $\pt$. 

The double-peak structure of the away-side correlation is consistent not only with conical emission, but also with other scenarios including large angle gluon radiation \cite{gluonrad} and jet ``deflection'' due to radial flow or the preferential selection of particles by the pathlength dependent energy loss mechanism~\cite{Hwa}. In order to distinguish between conical emission and other mechanisms, 3-particle azimuthal correlations are needed.

\section{Three-Particle Results and Discussions}

In 2- and 3-particle correlation analyses, a trigger particle at large $\pt$ is selected. Given a trigger particle, the event is composed of two parts: one directly correlated with the trigger (the so called ``dijet''), and the other not directly correlated (background). The background is {\em indirectly} correlated with the trigger via the reaction plane (flow correlation). The background is normalized to the 2-particle correlation signal by the common practice of ZYA1 or ZYAM (zero yield at 1 radian or minimum). Two combinatorial backgrounds are present in 3-particle correlation to a trigger particle: pairs of background particles and pairs of a correlated particle and a background particle. In the STAR results shown below, these backgrounds have been subtracted. The details of the 3-particle jet-correlation analysis is described in \cite{method}. 

Since the backgrounds are large in 3-particle correlations, it is critical to carefully construct the backgrounds. Analysis without careful background construction, for instance the 3-particle cumulant analysis~\cite{cumulant} where the background is blindly taken as the constant, average multiplicity density, can result in complex structures that are practically impossible to interpret~\cite{jetvscumu}.

\begin{figure}
\centering
\includegraphics[width=0.90\textwidth,bbllx=0,bblly=15,bburx=565,bbury=320]{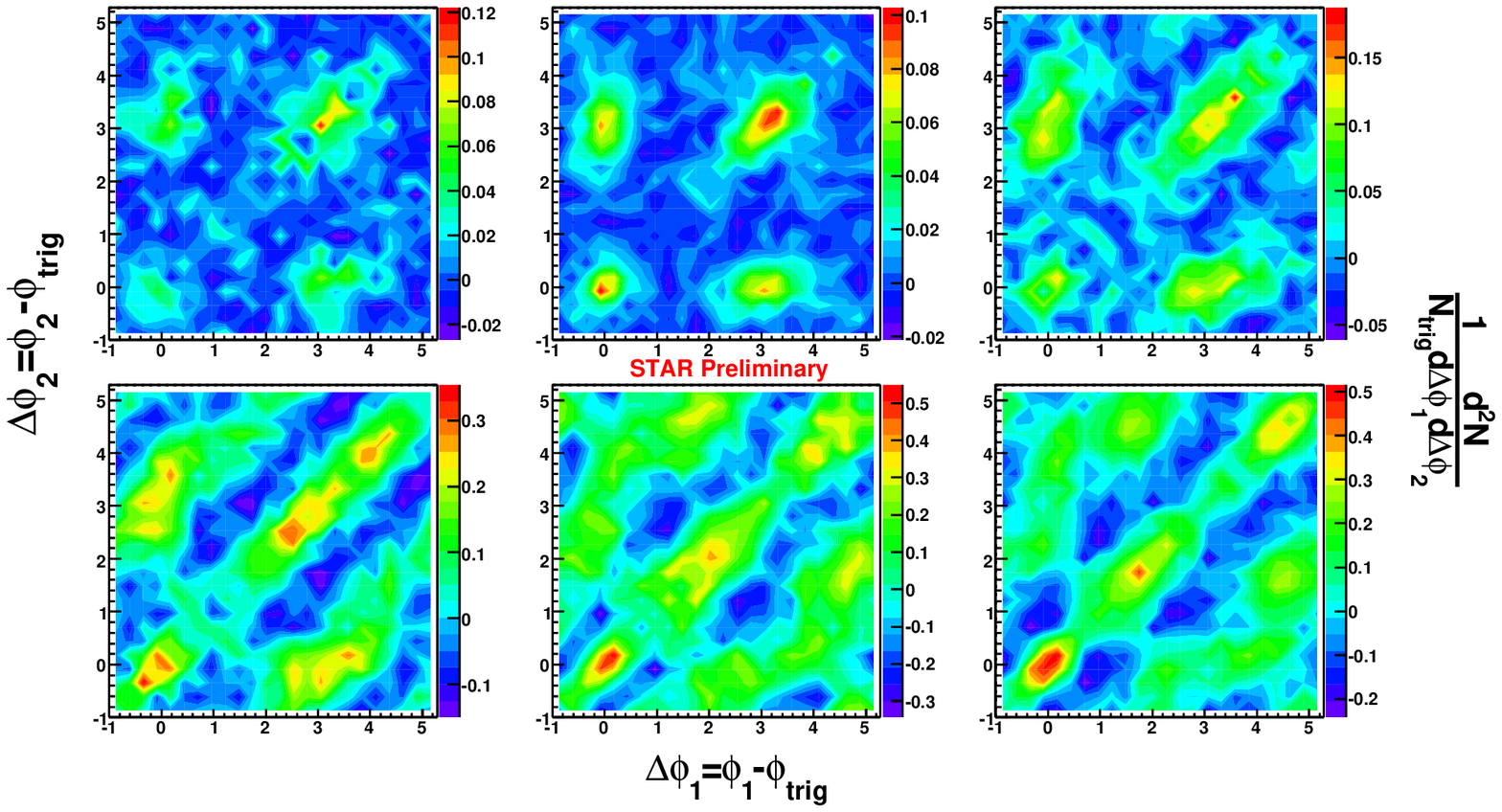}
\caption{(color online) Background subtracted 3-particle jet-like azimuthal correlations from STAR for {\it pp} (top left), d+Au (top middle), and Au+Au 50-80{\%} (top right), 30-50{\%} (bottom left), 10-30{\%} (bottom center), and ZDC triggered 0-12{\%} (bottom right). The figure is taken from~\cite{UleryHP}.}
\label{fig2}
\end{figure}

Figure~\ref{fig2} shows the 3-particle correlations between a trigger charged particle with $3<\pt^{trig}<4$ GeV/$c$ and two associated charged particles of $1<\pt<2$ GeV/$c$ measured by the STAR TPC~\cite{UleryHP}. The {\it pp}, d+Au and peripheral 50-80\% Au+Au results are similar.  Peaks are clearly visible for the near-side, the away-side and the two cases of one particle on the near-side and the other on the away-side.  The peak at ($\pi$,$\pi$) displays a diagonal elongation, consistent with $k_T$ broadening.  The additional broadening in Au+Au may be due to deflected jets.  The more central Au+Au collisions display off-diagonal structure, at about $\pi\pm1.3$ radian, that is consistent with conical emission.  The structure increases in magnitude with centrality and is quite clear in the high statistics 12\% central data~\cite{UleryHP}.

\begin{figure}[htbp]
\includegraphics[width=0.44\textwidth,bbllx=0,bblly=15,bburx=550,bbury=380]{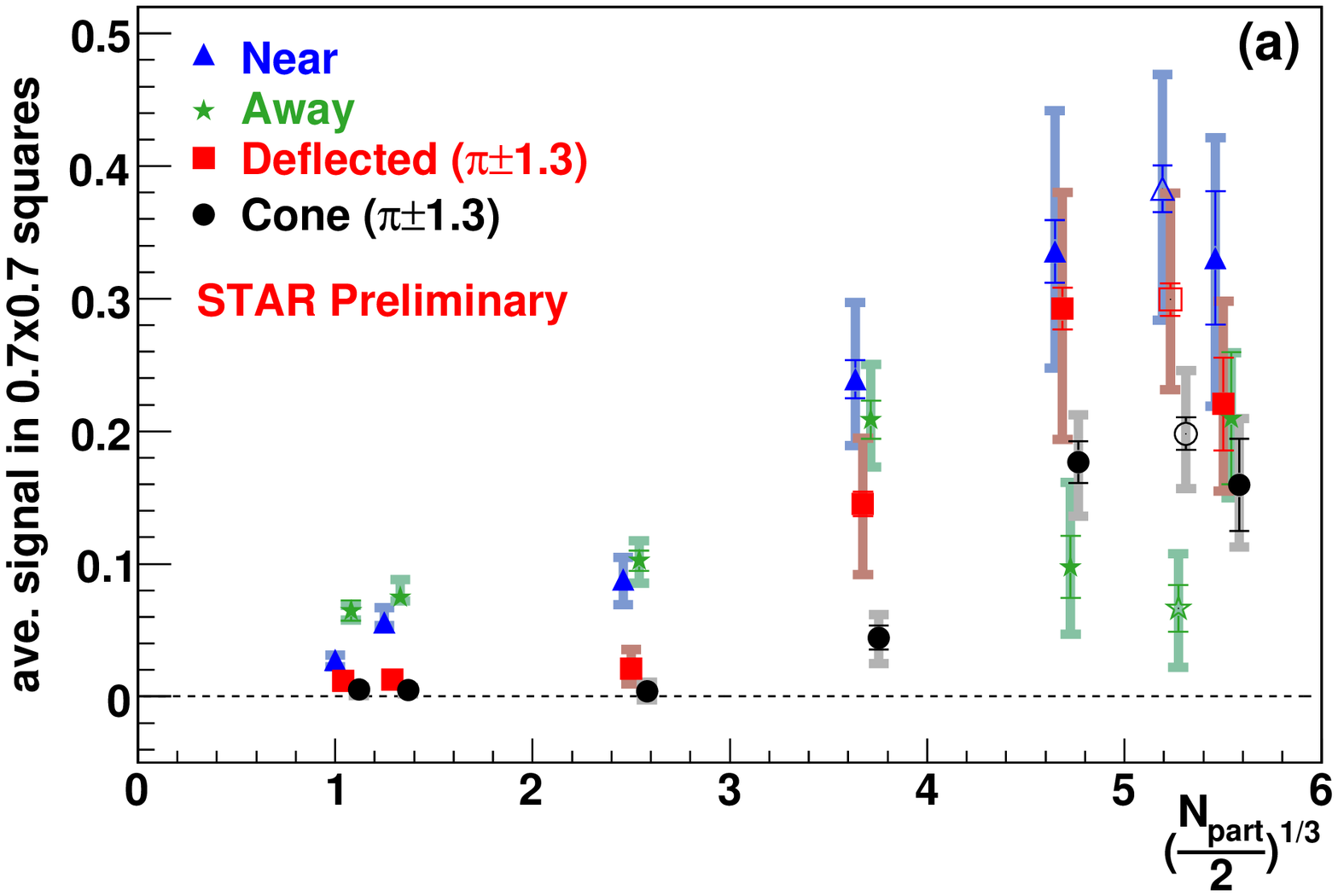}
\includegraphics[width=0.44\textwidth,bbllx=0,bblly=15,bburx=550,bbury=380]{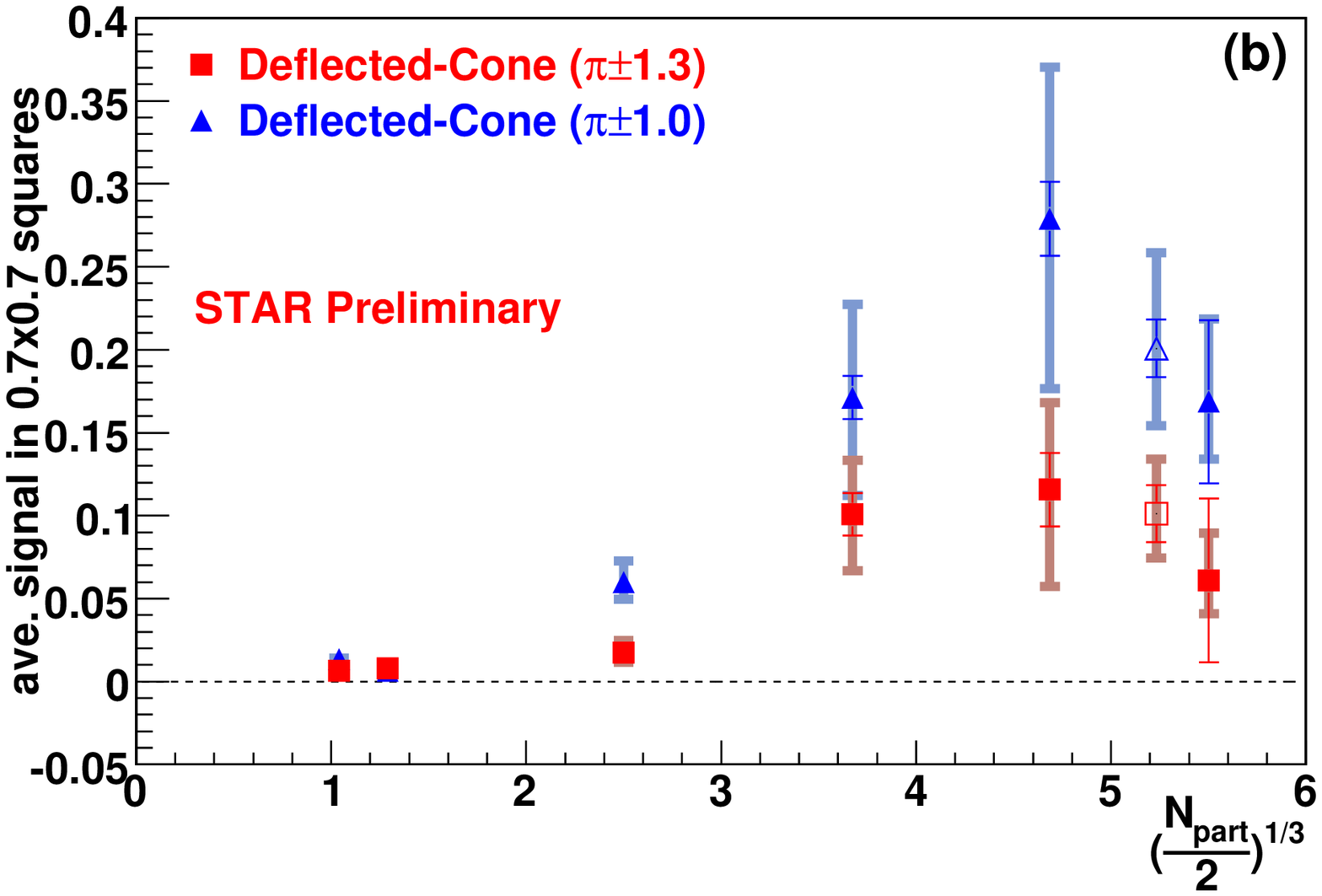}
\caption{(color online) (a) Average signals in 0.7 $\times$ 0.7 boxes at (0,0) (triangle), ($\pi$,$\pi$) (star), ($\pi\pm1.3$,$\pi\pm1.3$) (square), and ($\pi\pm1.3$,$\pi\mp1.3$) (circle).  (b) Differences between average signals, between ($\pi\pm1.3$,$\pi\pm1.3$) and ($\pi\pm1.3$,$\pi\mp1.3$) (square), and between ($\pi\pm1.0$,$\pi\pm1.0$) and ($\pi\pm1.0$,$\pi\mp1.0$) (triangle).  Solid error bars are statistical and shaded are systematic.  $N_{part}$ is the number of participants.  The ZDC 0-12\% points (open symbols) are shifted to the left for clarity. The figure is taken from~\cite{UleryHP}.}
\label{fig3}
\end{figure}

Figure~\ref{fig3}a shows the centrality dependence of the average signal strengths in different regions~\cite{UleryHP}. The off-diagonal signals (circle) increase with centrality and significantly deviate from zero in central Au+Au collisions. Figure~\ref{fig3}b shows the differences between on-diagonal signals, where both conical emission and deflected jets may contribute, and off-diagonal signals, where only conical emission contributes.  Since conical emission signals are of equal magnitude on-diagonal as off-diagonal, the difference may indicate the contribution from deflected jets.  The difference decreases with distance from ($\pi$,$\pi$).

The measured 3-particle jet-correlation structure in central Au+Au collisions is consistent with conical emission. The discrimination between Mach-cone shock waves and $\check{\rm C}$erenkov radiation needs future $\pt$-dependent studies. If the measured structure is indeed from Mach-cone shock waves, then it is possible to extract the conical emission angle, thereby the speed of sound of the medium created in these collisions. It is highly likely that the system evolves through different stages which the initially produced dijet probes: the partonic stage quark-gluon plasma, the mixed phase, and the hadronic stage. The measured conical emission is likely a net effect of all these stages; the extracted speed of sound is, therefore, an average over the evolution of the medium. The nature (or the equation of state) of the medium, however, needs careful investigations.

\section{CONCLUSION}

Broad, and for some kinematic regions even double-peaked, structures were observed on the away side of the 2-particle jet-correlations. The average transverse momentum of the away-side correlated hadrons is the lowest in the most collimated region. Mach-cone shock waves were proposed to explain the observations, however, other physics mechanisms cannot be ruled out without the knowledge of 3-particle azimuthal correlations. The 3-particle jet-correlation results from STAR are discussed. The central collision data show clear evidence of conical emission; they also indicate the presence of deflected jets. If Mach-cone shock waves are confirmed, further studies should be possible to extract the speed of sound (and the equation of state) of the medium, thereby providing crucial evidence for the creation of the quark-gluon plasma at RHIC.


\begin{theacknowledgments}
The author would like to thank Dr. Yiota Foka for the kind invitation. This work is supported by U.S. DOE under Grants DE-FG02-02ER41219 and DE-FG02-88ER40412.
\end{theacknowledgments}


\begin{thebibliography}{01}

\bibitem{Eloss}
R. Baier, D. Schiff, B.G. Zakharov, Annu. Rev. Nucl. Part. Sci. {\bf 50} (2000) 37;
X.-N. Wang and M. Gyulassy, Phys. Rev. Lett. {\bf 68} (1992) 1480.

\bibitem{wp}
J. Adams {\it et al.} (STAR Collaboration), Nucl. Phys. {\bf A757} (2005) 102;
K. Adcox {\it et al.} (PHENIX Collaboration), Nucl. Phys. {\bf A757} (2005) 184.

\bibitem{jetspectra}
J. Adams {\it et al.} (STAR Collaboration), Phys. Rev. Lett. {\bf95} (2005) 152301. 

\bibitem{dipPHENIX}
S.S. Adler {\it et al.} (PHENIX Collaboration), Phys. Rev. Lett. {\bf97} (2006) 052301.

\bibitem{awaympt}
F. Wang (STAR Collaboration), J. Phys. Conf. Ser. {\bf 27} (2005) 32 [nucl-ex/0508021];
F. Wang (STAR Collaboration), Nucl. Phys. {\bf A774} (2006) 129 [nucl-ex/0510068].

\bibitem{UleryQM05}
J.G. Ulery (STAR Collaboration), Nucl. Phys. {\bf A774} (2006) 581 [nucl-ex/0510055].

\bibitem{machcone}
H. Stoecker, Nucl. Phys. {\bf A750} (2005) 121;
J. Casalderrey-Solana, E. Shuryak and D. Teaney, J. Phys. Conf. Ser. {\bf27} (2005) 23.

\bibitem{Cerenkov}
I.M. Dremin, Nucl. Phys. {\bf A767} (2006) 233;
V. Koch, A. Majumder and X.-N. Wang, Phys. Rev. Lett. {\bf 96} (2006) 172302.

\bibitem{gluonrad}
I. Vitev, Phys. Lett. B {\bf 630} (2005) 78;
A.D. Polosa and C.A. Salgado, hep-ph/0607295.

\bibitem{Hwa}
R. Hwa, nucl-th/0609017.

\bibitem{method}
J.G. Ulery and F. Wang, nucl-ex/0609016.

\bibitem{cumulant}
C. Pruneau, nucl-ex/0608002.

\bibitem{jetvscumu}
J.G. Ulery and F. Wang, nucl-ex/0609017.

\bibitem{UleryHP}
J.G. Ulery (STAR Collaboration), nucl-ex/0609047.

\end{thebibliography}
\end{document}